# Power and Frequency Control of Induction Furnace Using Fuzzy Logic Controller

Behzad Sinafar
Department of Electrical and Computer Engineering ,
Sharif University of  Technology,
Tehran, Iran
Sinafar@ee.sharif.edu

Amir Rikhtegar Ghiasi
Faculty of Electrical and Computer Engineering
University of  Tabriz,
Tabriz, Iran
agiasi@tabrizu.ac.ir

*Abstract*—**This paper introduces a new method to control resonance frequency and output power of induction heating coil. Induction heating coil can be controlled by single phase sinusoidal pulse width modulation (SPWM) inverter .All electrical requirements beside magnetic permeability and resistivity variation for modeling  induction heating coil have been considered to make simulations practical .Control blocks using Fuzzy logic which control both active and reactive power have been designed .The system modeling and Fuzzy logic controllers are simulated in MATLAB/SIMULINK and FUZZY LOGIC toolbox .The results of the simulations  show the effectiveness and superiority  of this control system.**

*Keywords : Induction heating ,Fuzzy control ,Power control , SPWM , Resonance.*

## I. Introduction

With the development of power semiconductor devices, induction heating power supply has been more and more used for the heating process. Because of good heating efficiency, high production rate and clean working environments the induction heating process is widely used in industrial operating like metal hardening, preheating for forging operations, brazing and melting. But there are also some disadvantages, such as inertia, lag serious, nonlinear parameters and complex structure. As the heated object is unstable, induction heating power results in instability, inefficiency. Explicitly when a material is heated within induction furnace, its resistivity and magnetic permeability fluctuate, an increase in temperature leads to a rise in this material property. Resulting in the material's total resistance (R) increasing. This, in turn, causes a significant drop in the power drawn from the source. Meanwhile, as the temperature of a material approaches its Curie point, its magnetic permeability decreases, eventually dropping to unity at the Curie point. This leads to a reduction in inductance (L) as well as a change in resonance frequency. As a result, the active power reduces and an increase in reactive power descends the efficiency rate and power factor of load.

In order to make the inverter work always at a high power factor, the load must stay at resonant for whole heating process time. This requires an automatic phase control circuit. At present, induction heating power supply systems are generally use integrated phase locked loop (PLL), such as CD4046 is used to track the phase of inverter's output. However, these systems are complicated, difficult to calibrate, and sensitive to noise. Therefore, a simpler and more reliable way of implementing the system is suggested.

In this paper a simple and reliable way to determine resonance will be introduced and explained, which is derived from properties of resonant circuits. Hence, in this paper, we first measure active and reactive power by sampling output voltage of inverter and current of load series with compensator capacitance. Then a Fuzzy controller block is used to reduce reactive power and enables a variable induction heating load to be driven at its resonant frequency. A change in main switching frequency in order to track resonant frequency will affect active power transferred to load .consequently the main Fuzzy controller block will adjust the carrier frequency of PWM signals which results in the output power tracking the reference value. It should be noted that without using resonant controller the efficiency of system will decrease. Operating at resonance also has the advantage of ensuring reduced switching losses in the power source.

## II. SYSTEM CONFIGURATION

### A. The working principle of induction furnace

Medium Frequency Induction Furnace is mainly constituted by the furnace, inverter system, electrical control system, cooling system and etc; also the structure is more complicated. The main circuit topology of induction heating based on fuzzy control system is shown in Fig.1.

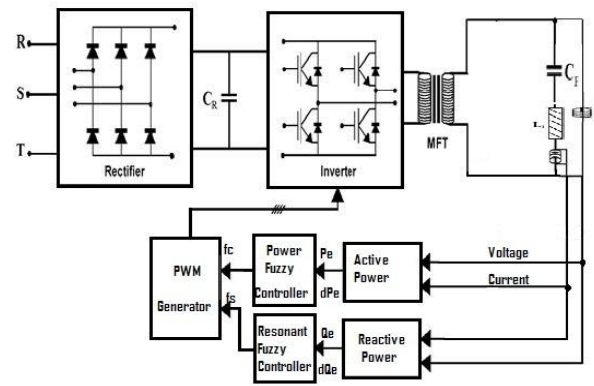

Figure 1. Block diagram of induction heating control system

As it's shown in the above figure, the plant system consists of three-phase rectifier, DC filter, single phase full bridge inverter and load series with compensator capacitance. In the induction furnace, energy is transferred directly from an induction coil into the material to be melted through the electromagnetic field produced by the induction coil. When an AC voltage is applied to the coil terminals, an alternating current passes through the coil winding. The current in each turn generates an electromagnetic field around it. This electromagnetic field induces current flowing around the surface of the load. This is called the skin effect or Kelvin effect. From this effect, one can easily infer that the heat energy converted from electric energy is concentrated on the skin depth (surface of the object).Between the heat energy transferred to load and output power of inverter, there is a relation as following equation:

$$P = \frac{cm}{0.24tn}T \quad (1)$$

P: output power transferred to load
T: the temperature of the furnace
c: the specific heat of the metallic charge
m: the weight of the metallic charge
t: heating time
n: the efficiency of the power

From the equation we can find that the temperature of furnace keeps a strict linear relation with the power output .Therefore, we can control the furnace temperature by regulating the output power exactly.

### B. Sinusoidal Pulse Width Modulation(SPWM) Inverter

The firing signals of switching gates re provided by PWM technique. Previously the main idea to control the power of induction furnace was using controlled rectifier and PLL blocks, meanwhile the PLL block adjusts the resonance frequency and controlled rectifier changes the output power of inverter by adjusting the amplitude of output voltage or using PWM generator by changing the width of gate pulses.

In this paper sinusoidal pulse width modulation (SPWM) technique will be used. The PWM inverter has been the main choice in power electronics for decades, because of its circuit simplicity and rugged control scheme. SPWM switching technique is commonly used in industrial applications.

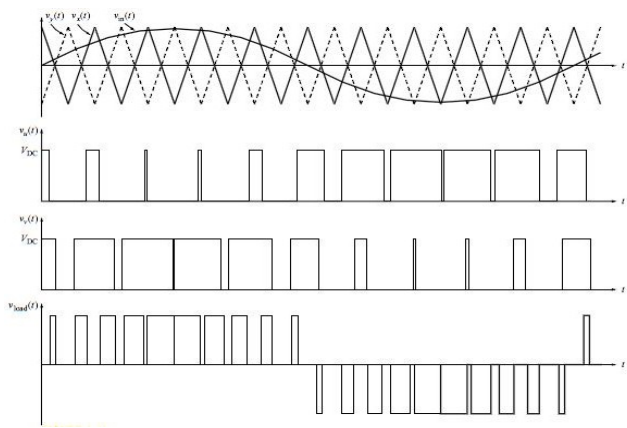

Figure 2. Sinusoidal pulse width modulation(SPWM)

SPWM techniques are characterized by constant amplitude pulses with different duty cycle for each period. The widths of these pulses are modulated to obtain inverter output voltage control The advantages of SPWM in comparison to PWM is that there are two degrees of freedom in SPWM which means by using this technique both resonance and power control will be provided and also the distortion factor and lower order harmonics will be reduced significantly.

The gating signals as shown in Fig.2 are generated by comparing a sinusoidal reference signal with a triangular carrier wave. The frequency of reference signal $f_s$ determines the inverter output frequency which should be kept equal to resonance frequency of the circuit. The number of pulses per half-cycle and width of each pulse depends on the carrier frequency $f_c$ . In this method the modulation index M which controls the amplitude of output voltage is considered constant.

### C. Induction Heating Load

The induction heating load comprises of a series resonant circuit, which is continuously driven at its natural resonant frequency by the inverter. It is evident that at a resonant frequency, the impedance of the load circuit is absolutely resistive. At this frequency the phase displacement between the driving voltage and current is equal to zero. This implies that maximum active power transfer is taking place and reactive power is equal to zero. These two rules are the bases of this paper to control power transferred to load and enables a variable induction heating load to be driven at its resonant frequency.

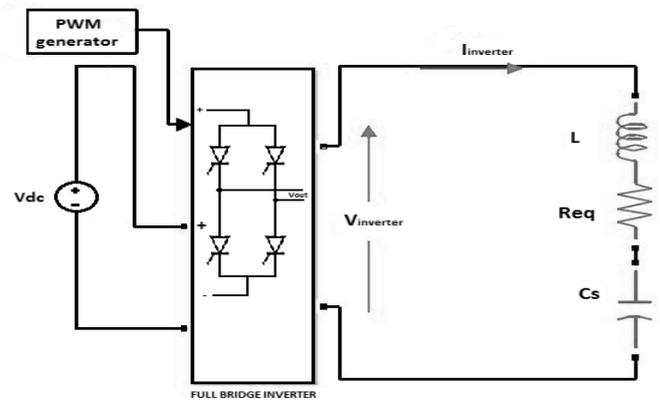

Figure 3.electrical circuit of series resonant inverter

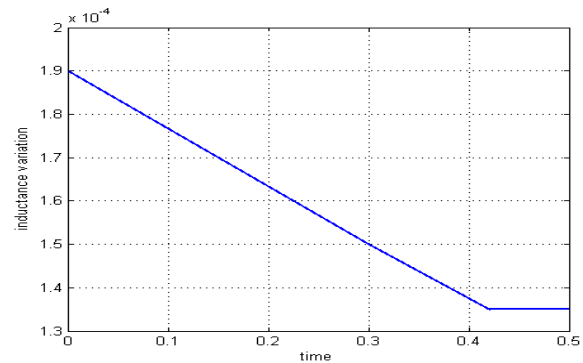

Figure 4.Inductance (L) variation due to material's thermal specifications and Curie point

Electrical equivalent circuit of load series with compensative capacitor is shown in Fig.3 .For simplicity the rectifier and input power sources are replaced with a dc power supply . The coil is simulated with a series connection of a resistor and an inductor. The initial values for resistor and inductance are R=21.90 mOhm and L=0.190 mH. As it's been explained in part I the induction heating load is time-varying parameters, non-linear structure and so on. Thermal variant parameters of load such as Magnetic permeability and resistivity of the materials experience large variations during heating in induction heating furnaces. Resulting in the equivalent resistance and inductance and also in output power. The variation of resistor may be neglected but the main problem will be caused by inductance variation which will ensure demolition of the resonance and that means existence of reactive power which will reduce efficiency and augment switching losses in the power source. The variation of inductance is shown in Fig.4.

## III. SUGGESTED CONTROL METHOD

### A. Control Block Structure

As mentioned in part I we face two problems with induction heating furnace. First, induction heating load must be driven at its resonant frequency in order to maximize efficiency and reduce problems caused by reactive power transferred to load. The main idea is that in any resonance circuit when the fundamental switching frequency is lower than resonance frequency the equal impedance is capacitive which means negative reactive power and when the fundamental switching frequency is higher than resonance frequency the equal impedance is inductive which means positive reactive power. These two concepts of resonance are the basis of reactive power control. Reactive power deviation ($Q_e$) along with its rate of change ($dQ_e$) are the inputs of resonance control block. Second, in order to achieve desired value of temperature which has a linear relation to output active power, main control block will be designed. Active power is compared to a reference power which is constant in this paper and power error ($P_e$) will be given. The inputs of this controller are power error ($P_e$) and its rate of change ($dP_e$). The output of this controller is frequency of triangular carrier signal which controls the sequence of pulses in each cycle of inverter output voltage. This frequency keep an inverse relation with active power transferred to load. By means of these two controllers, power and resonance frequency control of induction heating furnace will be achieved.

### B. Theory of Fuzzy Control

There are two fuzzy controllers in this paper for the induction heating system. Power controller is a typical block diagram of a fuzzy logic controller but the resonance controller has no reference input.

Most commercial fuzzy products are rule-based systems that receive current information in the feedback loop from the device as it operates and control the operation of a mechanical or other device. A fuzzy logic system has four blocks as shown in Fig. 5. Crisp input information from the device is converted into fuzzy values for each input fuzzy set with the fuzzification block. The universe of discourse of the input variables determines the required scaling for correct per-unit operation. The scaling is very important because the fuzzy system can be retrofitted with other devices or ranges of operation by just changing the scaling of the input and output. The decision-making-logic determines how the fuzzy logic operations are performed (Sup-Min inference), and together with the knowledge base determine the outputs of each fuzzy IF-THEN rules. Those are combined and converted to crispy values with the defuzzification block. The output crisp value can be calculated by the center of gravity or the weighted average.

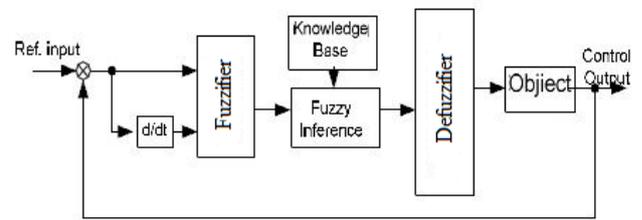

Figure 5.Fuzzy controller

### C. Design of Fuzzy Controller

In this paper, both controllers are two input and single output which is called two dimensional Fuzzy controller. All data transferred to or from Fuzzy controller are normalized. For the first controller (Reactive power) the input variables are $Q_e$ and $dQ_e$ and the output is $f_s$. the universe of discourse of input and output variables are divided into five fuzzy sets{NL,NS,Z,PS,PL}. TABLE I shows the Fuzzy rules which determines the relationship between fuzzy inputs and output. Also The membership functions of inputs and output are shown in Figs. 6. The statement which describes control rule can be written as "IF A is x AND B is y THEN C is z" where A,B are the fuzzy inputs and C is fuzzy output.

| $Q_e$ \ $dQ_e$ | NL | NS | Z | PS | PL |
|---|---|---|---|---|---|
| NL | PL | PL | PS | PS | Z |
| NS | PL | PS | PS | Z | NS |
| Z | PS | PS | Z | NS | NS |
| PS | PS | Z | NS | NS | NL |
| PL | Z | NS | NS | NL | NL |

Table I.fuzzy rules for resonance controller

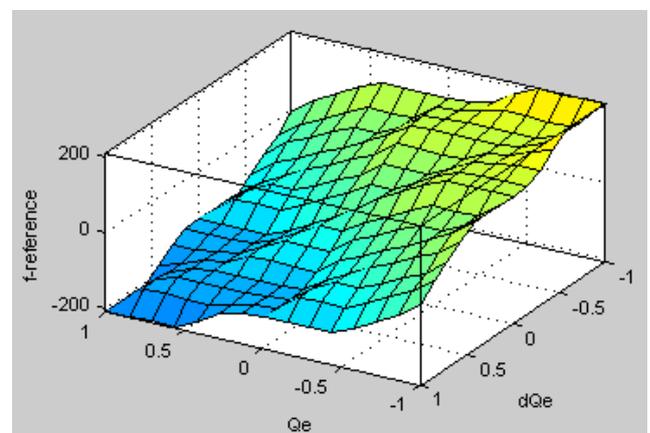

Figure 6. Surface observation of output reference frequency defuzzified by center of gravity

This controller has the advantage of driving the load at its resonance frequency but the issue that matter here is that as the output power is in relation with switching frequency, any change in resonance frequency will affect output power.

$$Z(f) = \frac{Rp}{1 + jQp(\frac{f}{fo} - \frac{fo}{f})} \quad (2)$$

where:
Rp = Equivalent resistance of the circuit as seen by the inverter,
Qp = Quality factor of the resonance circuit,
$fo$ = Natural resonant frequency of the circuit.

The main controller which will control the output power transferred to load is two dimensional Fuzzy controller with Pe and dPe as input and $f_s$ as output. The output of this controller will change the sequence of pulses and their width which will result in the active power transferred to load. As the power has a linear relation with temperature and is in inverse relation with frequency of carrier triangular signal, the fuzzy variables are divided into nine Fuzzy sets {NVL,NL,NM,NS,Z,PS,PM,PL,PVL}.TABLE II and Fig.8-10 respectively show the Fuzzy rules ,membership functions distribution of inputs and output and defuzzification of output.

| Pe \ dPe | NVL | NL | NM | NS | Z | PS | PM | PL | PVL |
|---|---|---|---|---|---|---|---|---|---|
| NVL | NVL | NVL | NVL | NVL | NVL | NL | NM | NS | Z |
| NL | NVL | NVL | NVL | NVL | NL | NM | NS | Z | PS |
| NM | NVL | NVL | NVL | NL | NM | NS | Z | PS | PM |
| NS | NVL | NVL | NL | NM | NS | Z | PS | PM | PL |
| Z | NVL | NL | NM | NS | Z | PS | PM | PL | PVL |
| PS | NL | NM | NS | Z | PS | PM | PL | PVL | PVL |
| PM | NM | NS | Z | PS | PM | PL | PVL | PVL | PVL |
| PL | NS | Z | PS | PM | PL | PVL | PVL | PVL | PVL |
| PVL | Z | PS | PM | PL | PVL | PVL | PVL | PVL | PVL |

TABLE II. FUZZY RULES FOR POWER CONTROLLER

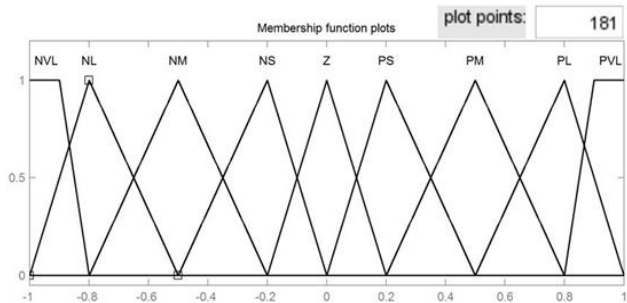

FIGURE 7. MEMBERSHIP DISTRIBUTION OF FUZZY INPUT VARIABLES (POWER ERROR AND POWER DIFFERENTIAL) FOR POWER FUZZY CONTROLLER

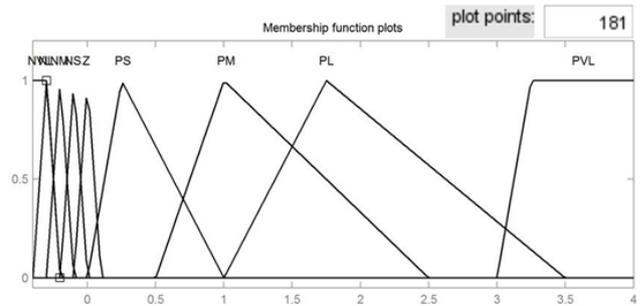

FIGURE 8. MEMBERSHIP DISTRIBUTION OF FUZZY OUTPUT VARIABLE (CARRIER FREQUENCY) FOR POWER FUZZY CONTROLLER

## IV. SIMULATION RESULTS

In this paper the following induction furnace parameters are used:

Fc=1000 Hz ,fs=250 Hz , Vdc =400 v ,Mindex=0.8

Using two Fuzzy controller blocks have the advantages of controlling output power beside reducing reactive power transferred to load which maximize the efficiency of system in comparison to other controllers.

Fig.11-12 show the output active and reactive power of open loop and closed loop system with Fuzzy controller. From the simulation results it can be seen that reactive power has been reduced significantly and input tracking for reference power has been performed successfully.

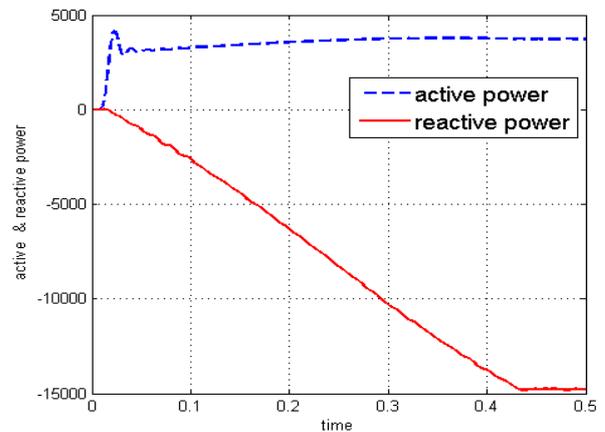

FIGURE 8. ACTIVE & REACTIVE POWER OF OPEN LOOP SYSTEM

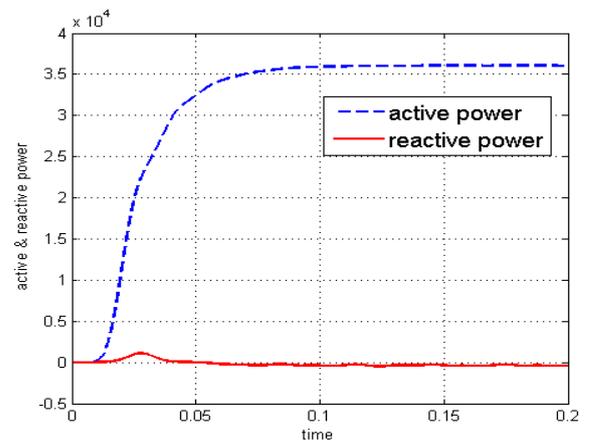

FIGURE 9. ACTIVE & REACTIVE POWER OF CLOSED LOOP SYSTEM WITH FUZZY CONTROLLER

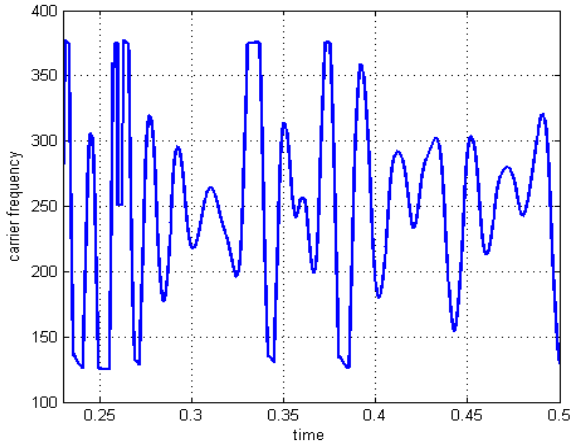

FIGURE 10. Frequncy of sinusoidal referenece signal used in SPWM block

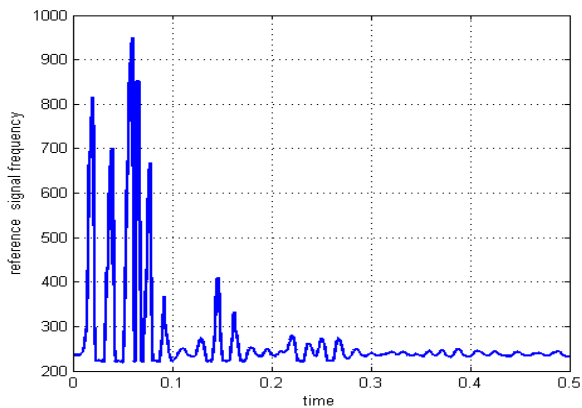

FIGURE 11. Frequency of carrier signal used in SPWM signal to track resonance

## V. COCLUSIONS

This paper presented a new method to control power and switching frequency of induction heating power supply using SPWM inverter. A fuzzy controller controlling SPWM carrier and output fundamental frequencies have been introduced. By means of the simulation carried out, an accurate and real model of induction heating load including both magnetic permeability and resistivity variation obtained. Meanwhile the load is variant. Without using PLL the FUZZY controller, by changing main switching frequency of the inverter, provides resonant state for load and by changing carrier frequency of the inverter provides SPWM signals for inverter such that output power of inverter will be adjusted to a desired value and reactive power will be reduced. By using this controller the power efficiency and control accuracy have been increased.

ACKNOWLEDMENTS

I would like to thank Dr .A .Giasi for valuable discussions and remarks and I wish to express my gratitude to his continuous encouragement.

I declare that this paper entitled "Power and Frequency Control of Induction Furnace Using Fuzzy Logic Controller "is the result of my own research except as cited in the references.